\documentclass[10pt,prl,preprintnumbers,floatfix,aps,nofootinbib,twocolumn,superscriptaddress]{revtex4} 

\usepackage[export]{adjustbox}
\usepackage[latin9]{inputenc}
\usepackage{graphicx}
\usepackage{amsmath}
\usepackage{amssymb}
\usepackage{amsfonts}
\usepackage{lscape}
\usepackage{hyperref}
\usepackage{url}
\usepackage{epsfig}
\usepackage{color}
\usepackage{bm}
\usepackage{tabularx}

\usepackage{url}
\usepackage[normalem]{ulem} 
\usepackage{latexsym}
\usepackage{float}
\usepackage{epsfig} 
\usepackage{verbatim}
\usepackage{enumerate,mdwlist}
\usepackage[titletoc]{appendix}
\usepackage{tikz} 
\graphicspath{{./Figures/}} 

\newcommand{\Msun}{M_\odot}

\newcommand{\OPBH}{\Omega_{\rm PBH}}
\newcommand{\DN}{\Delta N}

\newcommand{\lp}{\left(}
\newcommand{\rp}{\right)}
\newcommand{\lb}{\left[}
\newcommand{\rb}{\right]}

\newcommand{\ba}{\begin{eqnarray}}
\newcommand{\ea}{\end{eqnarray}}
\newcommand{\be}{\begin{equation}}
\newcommand{\ee}{\end{equation}}

\newcommand{\nn}{\nonumber}

\definecolor{grey}{rgb}{0.4,0.4,0.4}
\definecolor{dullmagenta}{rgb}{0.4,0,0.4}
\definecolor{darkblue}{rgb}{0,0,0.4}
\definecolor{midblue}{rgb}{0,0,0.5}
\definecolor{midred}{rgb}{0.5,0,0}
\definecolor{orange}{rgb}{1,0.5,0}
\definecolor{lightbrown}{rgb}{0.75,0.5,0.25}
\definecolor{tan}{cmyk}{0.14,0.42,0.56,0}
\definecolor{djunglegreen}{cmyk}{0.99,0,0.52,0}
\definecolor{lightgreen}{rgb}{0,1,0}
\definecolor{olivegreen}{cmyk}{0.64,0,0.95,0.40}
\definecolor{midgreen}{rgb}{0.0,0.675,0.0}
\definecolor{darkgreen}{rgb}{0,0.5,0}

\usepackage[all]{xy} 
\usepackage{amsfonts}

\makeatother

\begin{document}

\title{Primordial Black Hole production in Critical Higgs Inflation}

\author{Jose Mar\'ia Ezquiaga}
\email{jose.ezquiaga@uam.es}
\affiliation{Instituto de F\'isica Te\'orica UAM-CSIC, Universidad Auton\'oma de Madrid,
Cantoblanco, 28049 Madrid, Spain}
\affiliation{Berkeley Center for Cosmological Physics, University of California at Berkeley, 
Berkeley, California 94720, USA}

\author{Juan Garc\'ia-Bellido}
\email{juan.garciabellido@uam.es}
\affiliation{Instituto de F\'isica Te\'orica UAM-CSIC, Universidad Auton\'oma de Madrid,
Cantoblanco, 28049 Madrid, Spain}
\affiliation{CERN Theoretical Physics Department, 1211 Geneve, Switzerland}

\author{Ester Ruiz Morales}
\email{ester.ruiz.morales@upm.es}
\affiliation{Departamento de F\'isica, Universidad Polit\'ecnica de Madrid, 28012  Madrid, Spain}
\affiliation{CERN Theoretical Physics Department, 1211 Geneve, Switzerland}

\date{\today}

\begin{abstract}
Primordial Black Holes (PBH) arise naturally from high peaks in the curvature power spectrum of near-inflection-point single-field inflation, and could constitute today the dominant component of the dark matter in the universe. In this letter we explore the possibility that a broad spectrum of PBH is formed in models of Critical Higgs Inflation (CHI), where the near-inflection point is related to the critical value of the RGE running of both the Higgs self-coupling $\lambda(\mu)$ and its non-minimal coupling to gravity $\xi(\mu)$. We show that, for a wide range of model parameters, a half-domed-shaped peak in the matter spectrum arises at sufficiently small scales that it passes all the constraints from large scale structure observations. The predicted cosmic microwave background spectrum at large scales is in agreement with Planck 2015 data, and has a relatively large tensor-to-scalar ratio that may soon be detected by B-mode polarization experiments. Moreover, the wide peak in the power spectrum gives an approximately lognormal PBH distribution in the range of masses $0.01 - 100\,M_\odot$, which could explain the LIGO merger events, while passing all present PBH observational constraints. The stochastic background of gravitational waves coming from the unresolved black-hole-binary mergers could also be detected by LISA or PTA. Furthermore, the parameters of the CHI model are consistent, within $2\sigma$, with the measured Higgs parameters at the LHC and their running. Future measurements of the PBH mass spectrum could allow us to obtain complementary information about the Higgs couplings at energies well above the EW scale, and thus constrain new physics beyond the Standard Model. 
\end{abstract}
\maketitle

{\it Introduction}. The first direct detection of gravitational waves (GWs) by LIGO has initiated a new era of astronomy~\cite{LIGO} and opened the possibility to test the nature of dark matter, specially if its dominant component is primordial black holes (PBH)~\cite{PBH}. These massive black holes could arise in the early universe from the gravitational collapse of matter/radiation on large-amplitude curvature fluctuations generated during inflation~\cite{GBLW,CGB}. All that is required is a super-slow-roll period (i.e. a plateau feature in the potential) during which the inflaton quantum fluctuations get amplified and produce a peak in the spatial curvature power spectrum~\cite{single}. The mass and spin distribution of the generated PBH then depends on the details of the inflationary dynamics and their subsequent evolution during the radiation and matter eras. Its detection and characterization by LIGO, VIRGO and future GW detectors will allow us to open a new window into the physics of the early universe.

The nature of the inflaton field responsible for the initial acceleration of the universe is still unknown. Observations of the temperature and polarization anisotropies in the cosmic microwave background (CMB) suggests a special inflaton dynamics, dominated by a flat plateau on large scales~\cite{Planck}. Such type of potentials arise naturally in models of Higgs Inflation~\cite{HI}, where the scalar field responsible for inflation is the Higgs boson of the Standard Model (SM) of Particle Physics, with its usual couplings to ordinary matter (gauge fields, quarks and leptons), plus a new non-minimal coupling $\xi$ to gravity. This economical scenario not only passes all solar system and CMB observational constraints, but also predicts a small tensor-to-scalar ratio and a large reheating temperature~\cite{GBFR}.

Recent measurements of $\alpha_s$ and $m_{\rm top}$ hint at the possibility, envisioned by Froggatt and Nielsen in 1979~\cite{Froggatt:1978nt}, that the running of the Higgs self-coupling to large energy scales, via the renormalization group equations (RGE), could lead to a critical point $\phi=\mu$, with $\lambda(\mu) = \beta_\lambda(\mu)=0$, where $\lambda(\phi)$ has a minimum~\cite{Buttazzo:2013uya}. This scenario was explored in the context of critical Higgs inflation in Ref.~\cite{CHI}, generating the right amplitude of CMB anisotropies with a relatively small $\xi$ coupling. In this paper, we extend the analysis of~\cite{CHI} taking into account also the running of the $\xi$ coupling. This extra feature in the inflationary potential can induce a brief plateau of super-slow-roll conditions at scales much smaller than those of the CMB, giving rise to a peak in the matter power spectrum.

\begin{figure*}[!t]
\centering
\includegraphics[width = 0.47\textwidth]{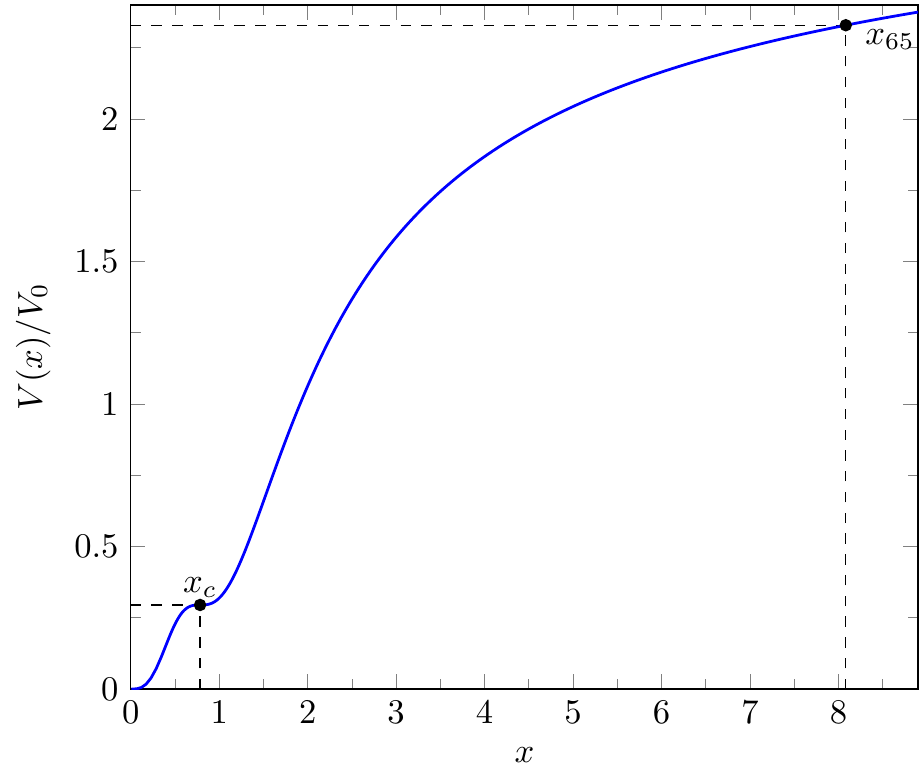}
\includegraphics[width =0.49 \textwidth]{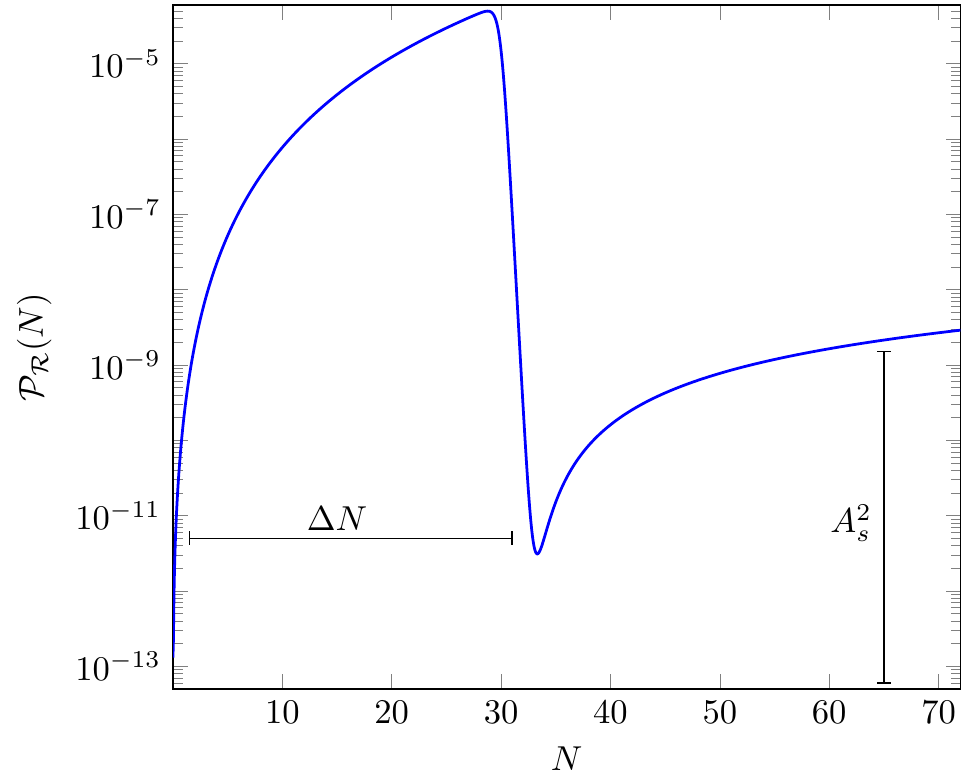}
\includegraphics[width =0.47\textwidth]{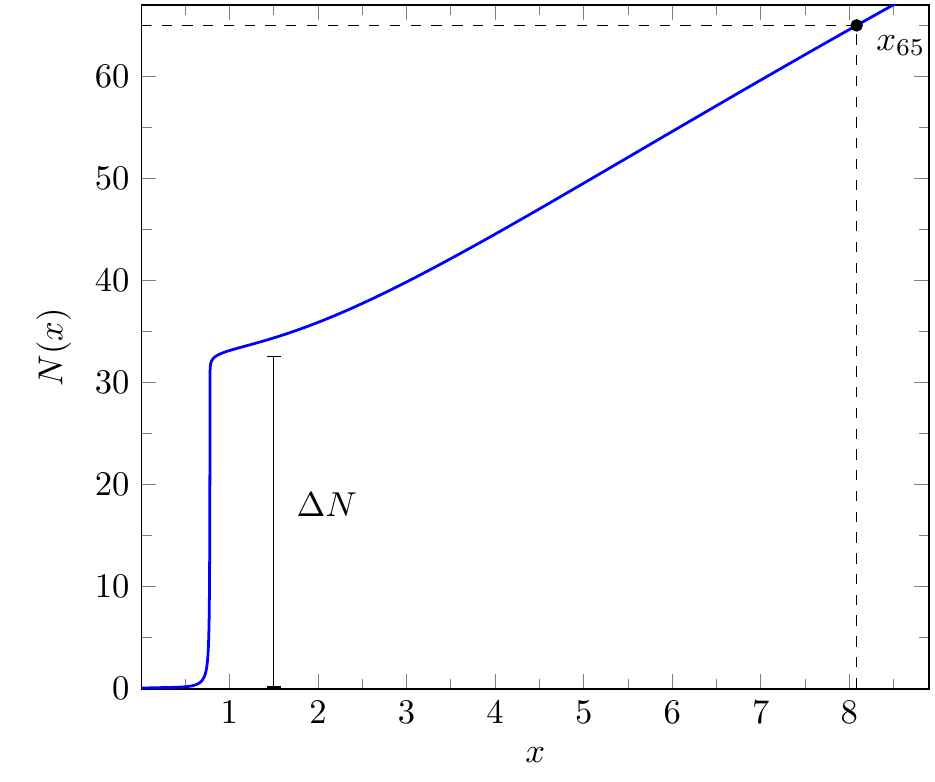}
\includegraphics[width = 0.49\textwidth]{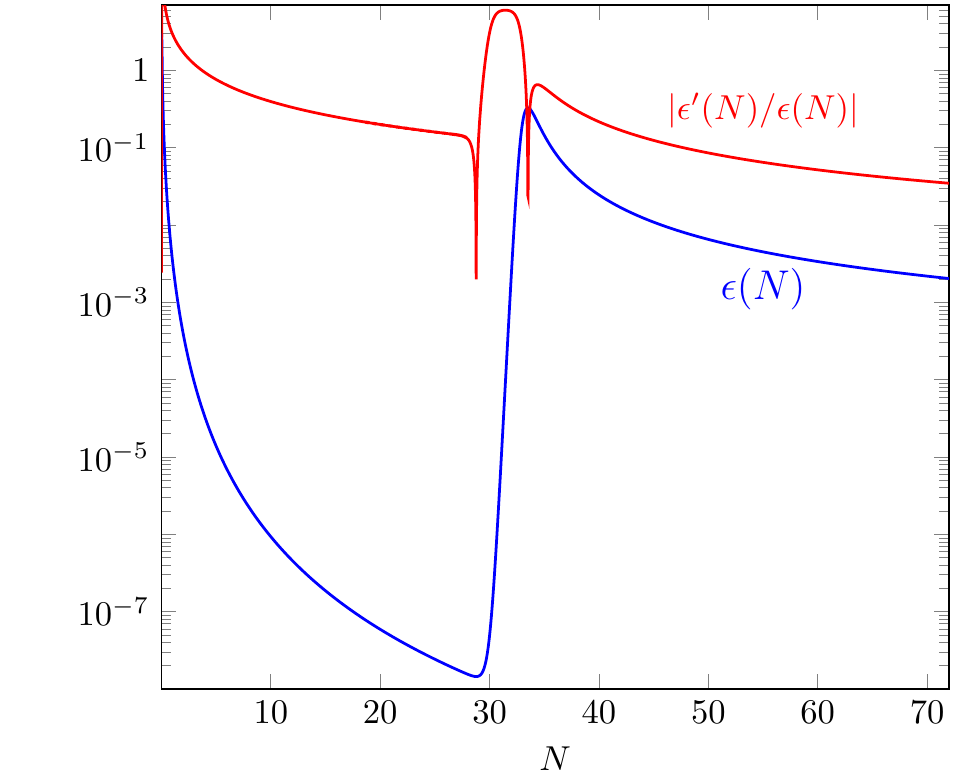}
\caption{\emph{Top panels:} the Critical Higgs Inflation potential (left) and its curvature power spectrum ${\cal P}_{_{\cal R}}(N)$ (right). The large and broad \emph{half-dome} peak at small scales ($N<\DN$) is responsible for PBH production over a wide range of masses. \emph{Bottom panels:} evolution of the number of $e$-folds (left) and the slow-roll parameters (right) for the exact equations of motion.}
\label{fig:PRN}
\end{figure*}
%

\begin{figure*}[!t]
\vspace{-10pt}
\centering
\includegraphics[width = 0.50\textwidth,valign=t]{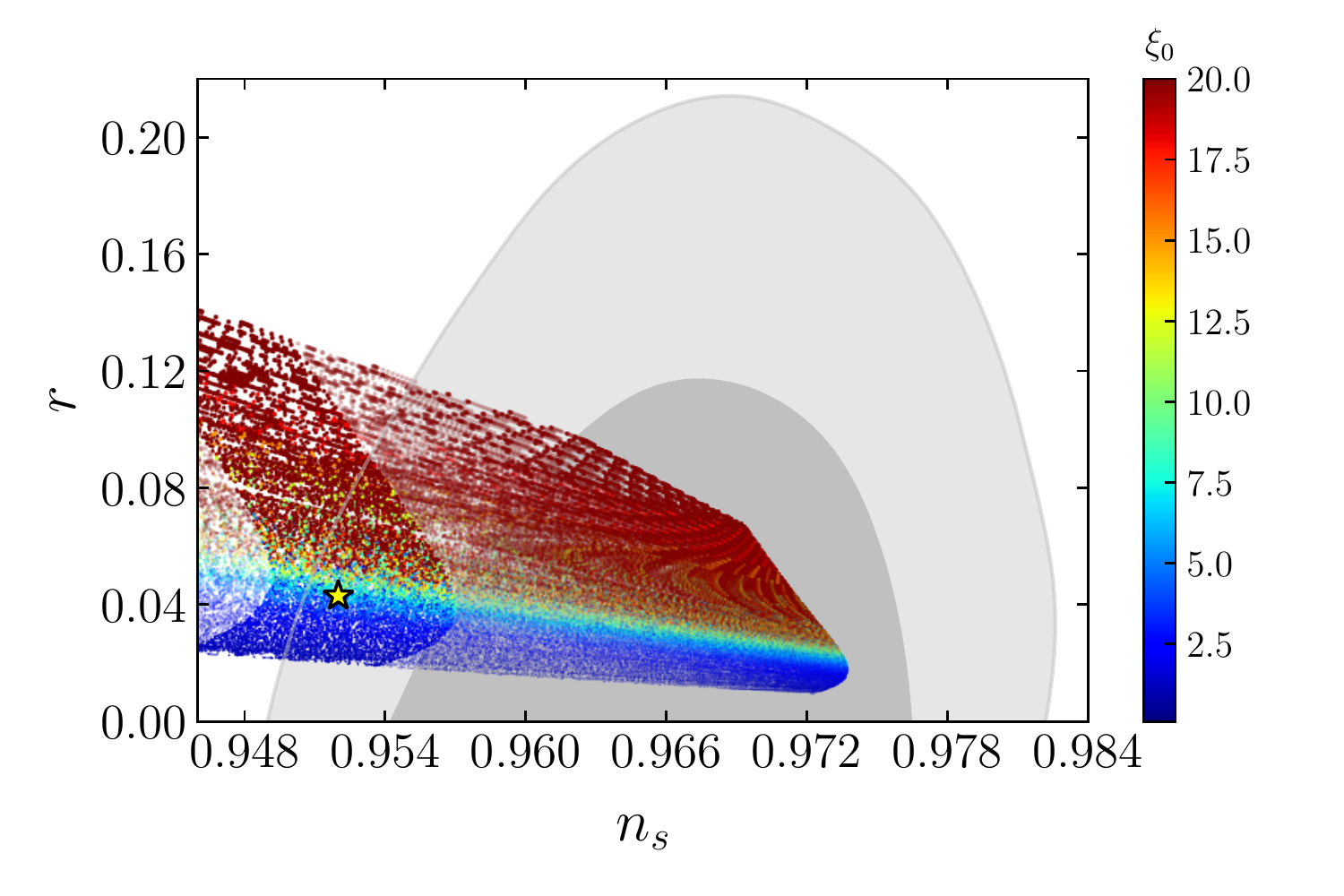}
\includegraphics[width = 0.48\textwidth,valign=t]{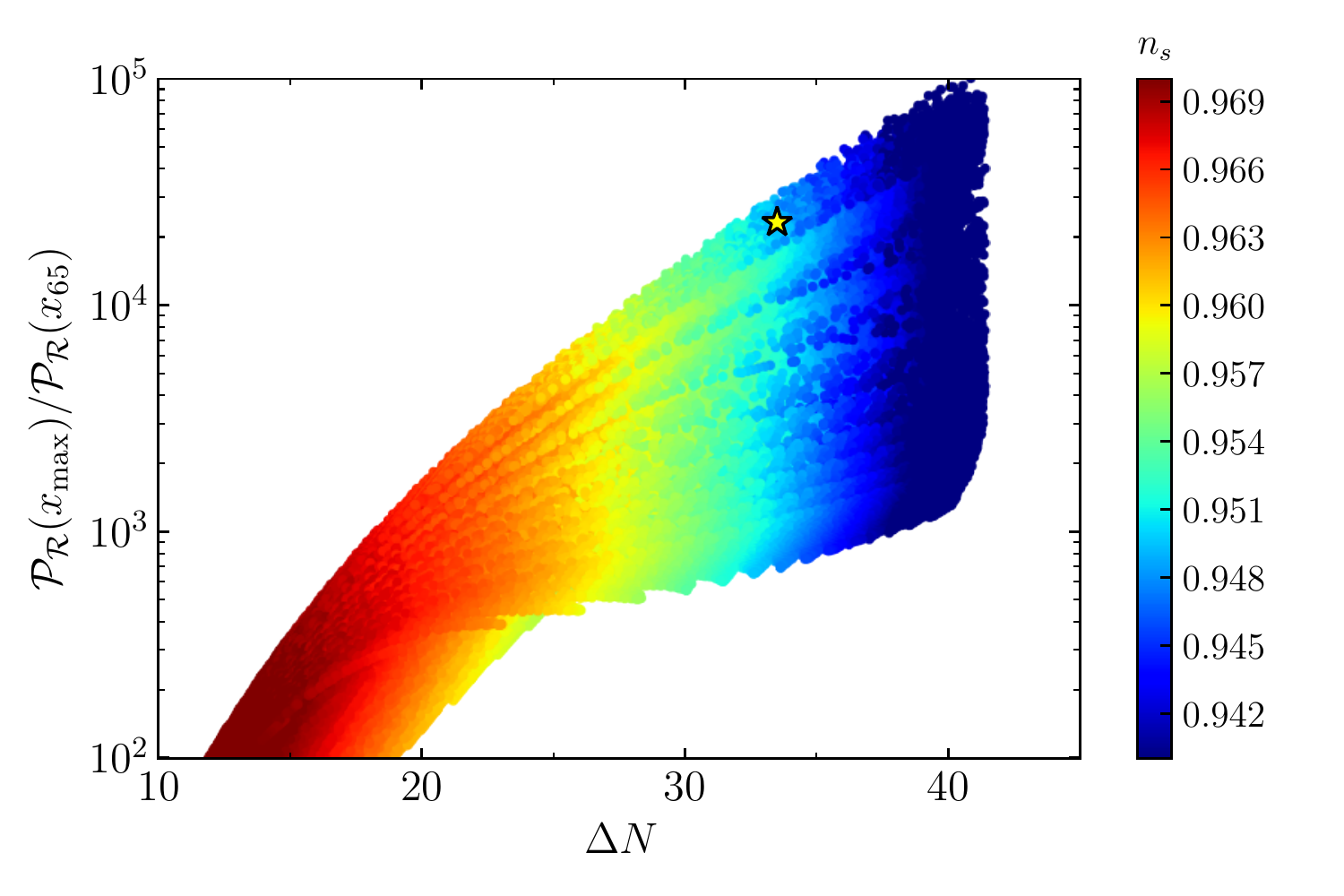}
\vspace{-10pt}
\caption{\emph{Left-panel:} $(n_s,\,r)$-plane of CHI. The region with denser color corresponds to $\DN\in(30,35)$ and the contours represents the 1 and $2\sigma$ Planck constraints for models with variable $n_s$, $dn_s/d\ln k$ and $r$, obtained from the Planck Legacy Archive. \emph{Right panel:} height of the peak as a function of its width. In both cases, the star corresponds to the reference parameter choice with $n_s=0.952$, $r=0.043$, $\DN=33.5$ and ${\cal P}_{_{\cal R}}(x_{\mathrm{max}})/{\cal P}_{_{\cal R}}(x_{65})= 2.3\times 10^{4}$ and the other points are all within $\beta\in(0.1 - 9)\times10^{-4}$ and $\DN\in(10,45)$.}
\vspace{-10pt}
\label{fig:nsr}
\end{figure*}

\vspace{2mm}
{\it Critical Higgs Inflation, CMB and Particle Physics}. In this letter, we explore this critical Higgs scenario, taking into account {\em both} the RGE running of the Higgs self-coupling and its non-minimal coupling to gravity~\cite{Herranen:2015ima}. The action of the Higgs-inflaton model is given by 
\be
S\!=\!\!\int \!d^{4}x\sqrt{g} \lb \lp\frac{1}{2\kappa^2}
+\frac{\xi(\phi)}{2}\phi^{2}\rp\!R -\frac{1}{2}(\partial\phi)^{2}-\frac{1}{4}\lambda(\phi)\phi^{4}\rb\,,
\ee
where $\kappa^2\equiv8\pi G=1/M^2_{\rm Pl}$, and we have expanded the running of the couplings around the critical point,
$\phi=\mu$, as
\ba
\lambda(\phi)&=&\lambda_0+b_\lambda \ln^{2}\lp\phi/\mu\rp\,, \label{eq:RGEl} \\[2mm] \label{eq:RGExi} 
\xi(\phi)&=&\xi_0+b_\xi \ln\lp\phi/\mu\rp\,.
\ea

After a standard metric and scalar field redefinitions, 
\ba
&& g_{\mu\nu} \to \left(1+\xi(\phi)\phi^2\right)\,g_{\mu\nu}\,,\\[2mm]
&&\hspace*{-2mm} 
\hspace{-10pt}\phi \to \varphi=\hspace{-2pt}\int\hspace{-2pt} d\phi \frac{\sqrt{1+\phi^2(\xi(\phi)+6(\xi(\phi)+\phi\xi(\phi)'/2)^2)}}
{1+\xi(\phi)\phi^2}\,
\ea
the effective inflationary potential becomes
\be
V(x)=\frac{V_0\,(1+a\,\ln^{2}x)\,x^{4}}{(1+c\,(1+b\,\ln x)\,x^{2})^{2}}\,,
\label{eq:potential}
\ee
with $V_0 = \lambda_0\mu^4/4$, $a=b_\lambda/\lambda_0$, 
$b=b_\xi/\xi_0$ and $c=\xi_0\,\kappa^2\mu^2$. The potential has a flat plateau 
at large values of the field $x=\phi/\mu$, see top-left panel of Fig.~\ref{fig:PRN}, where
\be
V_\infty \equiv V(x\gg x_c) \simeq V_0 \,\frac{a}{(b\,c)^2} = \frac{1}{4\kappa^4}\,\frac{b_\lambda}{b_\xi^2} \ll M_{\rm P}^4\,.
\ee
Note that the small value of $H^2_{\rm inf} = \kappa^2 V_\infty/3 \ll M^2_{\rm P}$ is determined in this model by the RGE running of the SM Higgs couplings $\lambda$~and~$\xi$. The potential also has a short secondary plateau around the critical point, where the inflaton-Higgs slows down and induces a large peak in the curvature power spectrum.  This second plateau is induced by a near-inflection point at $x=x_c$, where $V'(x_c) \simeq 0,\ V''(x_c) \simeq 0$. As a consequence, the number of $e$-folds has a sharp jump $\Delta N$ at that point, cf. bottom-left panel of Fig.~\ref{fig:PRN}, plus a slow rise towards larger field values, corresponding to CMB scales.

This behavior is very similar to the one discussed in Ref.~\cite{single}. Following this reference, we have computed the exact inflationary evolution. One should notice that, although the slow-roll parameter $\epsilon(N)=\kappa^2\varphi'(N)^2/2$ is always smaller than one, its variation $\epsilon'(N)/\epsilon(N)$ can be larger around the near inflection point~\cite{SRA}. Still, for a large set of the CHI parameter space, the inflaton slows down around $x_c$, producing a broad peak in the spectrum, but keeps enough inertia to cross the near-inflection point and continue rolling down the potential towards the end of inflation in just a few e-folds. This is exemplified in the bottom-right panel of Fig.~\ref{fig:PRN}. Thus, CHI can produce a successful inflation with a characteristic \emph{half-dome} peak in the spectrum at small scales.

We chose to explore the predictions of the model in terms of the height and width of the peak in the power spectrum, see top-right panel of Fig.~\ref{fig:PRN}. The height of the peak relative to the amplitude at CMB scales ($A_s^2$) is controlled by the closeness of $x_c$ to a true inflection point, $V'(x_c)=V''(x_c)=0$. The width of the peak is determined by the jump in the number of $e$-folds, $\Delta N$. There will be a true inflection point at $x_c$ if
\ba
&&a(x_c,\,c)=\frac{4}{1+c\, x^2 _c +2\ln x_c -4\ln^2 x_c }\,, \label{eq:a}\\[1mm]
&&b(x_c,\,c)=\frac{2(1+c\, x_c ^2+4\ln x_c + 2c\, x^2_c \ln x_c)}
{c\, x_c ^2(1+c\, x^2_c +2\ln x_c -4\ln^2 x_c)} \label{eq:b}\,.
\ea
Thus, a near-inflection point can be characterized by $a\rightarrow a(x_c,\,c)$ and $b\rightarrow (1-\beta)\,b(x_c,\,c)$. Then, the relative height of the peak will be inversely proportional to $\beta$ and will increase with the width $\Delta N$. We explore the $(\beta,\,\xi_0,\,x_c,\,c)$ parameter space searching for power spectra consistent with the latest CMB constraints and producing a sizeable peak at $x_c$. The value of $\lambda_0$ is chosen to match to the observed CMB amplitude $A_s^2$. Therefore, 
there is a one-to-one correspondence between each point of the viable parameter space, and the parameters of the potential via Eqs.~(\ref{eq:a}$-$\ref{eq:b}).

We have studied the main CMB observables (the scalar spectral index $n_s$, its running, $\alpha_s=d n_s/d\ln k$, and the tensor-to-scalar ratio $r$), as a function of $(x_c,\,c)$, for different heights and widths. We find that, for each pair ($\beta,\,\Delta N$), there are many choices of $(x_c,\,c)$ that give rise to valid cosmologies. In particular, we have chosen as reference point in parameter space,
\be\label{eq:set}
\beta=10^{-5}\,,\hspace{2mm}
\Delta N=33.5\,,\hspace{2mm}
x_c = 0.784\,,\hspace{2mm}
c= 0.77\,,
\ee
which give the CMB parameters
\be
n_s=0.952\,, \ r=0.043\,, \ \alpha_s=-0.0017\,,
\ee 
perfectly within the $2\sigma$ limits of Planck 2015~\cite{Planck}.

We present in Fig.~\ref{fig:nsr}a the predictions of our model for a range of parameters in the $(n_s,\,r)$-plane for $\beta\in(0.1 - 9)\times10^{-4}$, and $\DN\in(10,45)$, together with the 1 and $2\sigma$ constraints from CMB anisotropies, as measured by Planck 2015, shown by the grey contours. We show in color the values of the non-minimal coupling $\xi_0$ in the $(n_s,\,r)$-plane. The region with denser color represents cases with $\DN\in(30,35)$, which produce a sufficiently large peak in the power spectrum at small scales to later give rise to PBH through gravitational collapse upon reentry~\cite{GBLW}. This region tends to give low spectral index, $n_s<0.956$, and large tensor-to-scalar ratios, $r>0.019$, while cases with lower 
$\DN$ display a better fit to Planck data but cannot generate significant populations of PBHs. In the right panel of Fig.~\ref{fig:nsr} we show the ratio ${\cal P}_{_{\cal R}}(x_{\mathrm{max}})/{\cal P}_{_{\cal R}}(x_{65})$ of the amplitude of the fluctuations at its maximum, $x_{\mathrm{max}}$, over the amplitude at the inflationary plateau, $x_{65}$, as a function of $\DN$. The color code indicates the spectral tilt $n_s$ for each particular case. This figure shows that significantly large ratios can only be obtained for large values of $\DN$, which are also associated with lower values of $n_s$.

The reference point (\ref{eq:set}) corresponds to the model parameters
\be
\begin{split}
\hspace*{-1.0cm}&\lambda_0=2.23\times10^{-7}\,, \
\xi_0=7.55\,, \hspace*{3mm} \kappa^2\mu^2=0.102\,, \\[2mm]
\hspace*{-1.0cm}& b_\lambda=1.2\times10^{-6}\,, \hspace*{3mm}
b_\xi=11.5\,.
\end{split}
\ee
In order to have a large PBH production and a good agreement with the CMB constraints, the allowed range of CHI parameters can be enlarged to 
$\lambda_0\sim(0.01-8)\times10^{-7}$, $\xi_0\sim(0.5-15)$, $\kappa^2\mu^2\sim(0.05-1.2)$, $b_\lambda\sim(0.008-4)\times10^{-6}$ and $b_\xi\sim(1-18)$, for $\DN\in(30,35)$. The question arises whether these values, corresponding to the model parameters at the critical scale $\mu$, are consistent with the values of the SM parameters at the EW scale. Given the latest values of $m_{\rm top}$ and $\alpha_s$, the values of $\lambda_0$ and $b_\lambda$ that we consider
for the Higgs quartic coupling, are consistent, within 2$\sigma$, with the existence of a critical point $\beta_\lambda(\mu)=\lambda(\mu)=0$ around scales $\mu \sim 10^{17} - 10^{18}$ GeV, via the RGE equations of the SM. On the other hand, the non-minimal coupling of the Higgs to gravity is still unknown, but is a natural consequence of quantum field theory in curved space time~\cite{Herranen:2015ima}. It has been argued that, to avoid contribution from higher-order effective operators to the Higgs potential, the coupling $\xi$ should be sufficiently small~\cite{Barbon:2009ya}. In our model, the inclusion of the running of $\xi$, in Eq.~(\ref{eq:RGExi}), gives reasonably small values for these parameters.

Future measurements of the PBH mass spectrum will allow us to obtain complementary information about the SM couplings of the Higgs at high energy scales, which allows one to have a large lever arm for the RGE running of these couplings from the EW scale to almost the Planck scale. A detailed analysis of the compatibility of these coefficients with the predictions of the SM non-minimally coupled to gravity requires further work, possibly with the inclusion of threshold corrections, see~\cite{Bezrukov:2017dyv}. 

It is also interesting to note that this CHI scenario predicts an amplitude of tensor modes that lies within the target range of present and next-generation B-mode experiments~\cite{Bmode}. Moreover, the large amplitude of curvature fluctuations a few $e$-folds before the end of inflation, see Fig.~\ref{fig:PRN}, may induce a significantly inhomogeneous reheating upon reentry, which could have important consequences for the reheating temperature and possibly also for the production of PBH and gravitational waves at preheating, see e.g.~\cite{GBF}. In particular, we find that the energy density at the end of inflation is $\rho_{\rm end}=4\times10^{63}$ GeV$^{4}$ and the estimated reheating temperature (for $g_*=106.75$), $T_{\rm rh}=3.2\times10^{15}$ GeV, is relatively high, justifying our choice of $N=65$ e-folds of inflation.

\vspace{2mm}
{\it Production of PBHs and DM}. We use the Press-Schechter formalism of gravitational collapse to compute the probability that a given horizon-sized volume forms a PBH when a large curvature fluctuation, $\zeta > \zeta_c$, reenters the horizon during the radiation era~\cite{Harada:2013epa}, and not even radiation pressure can prevent collapse, as described in Ref.~\cite{single}. Thus, the fraction of PBHs at formation can be computed from~\cite{CGB} 
\be
\beta_f(M)=\frac{\rho_{\rm PBH}}{\rho_{\rm tot}}(t_f)=
{\rm erfc}\lp\frac{\zeta_c}{\sqrt{2{\cal P}_{\cal R}}}\rp\,.
\ee
The mass of the PBH at formation is essentially given (within an order-one efficiency factor $\gamma$) by the total mass within the horizon at the time of reentry, i.e. $M_{\rm PBH}\sim \gamma M^2_{\rm Pl}\,e^{2N}/2H_{\rm inf}$. In our case, for the large and wide peak in ${\cal P}_{\cal R}(k)$ at small scales, one finds an approximate lognormal distribution of masses for PBH, 
\be
P(M) = \frac{A\,\mu}{M\,\sqrt{2\pi\sigma^2}}\exp\left(-\frac{\ln^2(M/\mu)}{2\sigma^2}\right)\,,
\ee
with a sharp drop at high masses due to the half-dome shape of the peak, see Fig.~\ref{fig:PRN}. This characteristic shape also shifts the peak of the mass spectrum to higher values since the PBH mass exponentially depends on the number of $e$-folds at reentry. The distribution of PBHs at equality is fully characterized by the physics of inflation through ${\cal P}_{\cal R}(k)$; its evolution during radiation domination, which linearly increases with the scale factor $\beta_{\rm eq}(M)=\beta_f(M)\cdot a_{\rm eq}/a(t_{_M})$; and the evaporation due to Hawking radiation, which erases the lightest PBHs. We find that for the range of $\DN\in(30-35)$, PBHs can constitute the total DM at equality, i.e. $\OPBH^{\rm eq}=\int\beta_{\rm eq}(M)\,d\ln M=0.42$, within the uncertainty range of $\zeta_c \sim (0.05-1)$~\cite{Harada:2013epa}. For the reference point in parameter space that we have chosen, we use $\zeta_c = 0.052$ and $\gamma = 0.4$. Here we do not consider any quantum diffusion during inflation \cite{Pattison:2017mbe} or a non-linear growth in mass before equality. These effects might increase the abundance of PBHs at equality $\OPBH^{\rm eq}$.

\begin{figure}[!t]
\hspace{-20pt}
	\centering
        \includegraphics[width=0.51\textwidth]{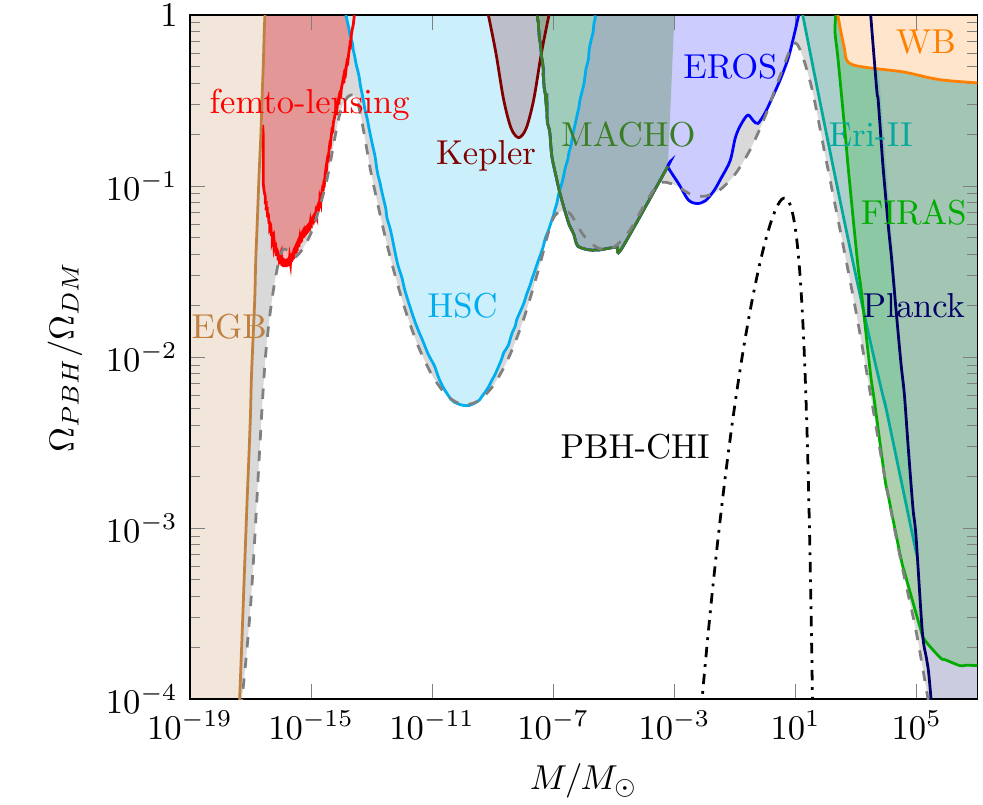}
     \centering
     \vspace{-10pt}
  \caption{Present constraints on PBH from Extragalactic Gamma Background (EGB), femto-lensing of GRB, micro-lensing (HSC, Kepler, MACHO and EROS), Wide Binaries (WB), Eridanus II (Eri-II) and the CMB (FIRAS and Planck). See Refs.~\cite{JGB,Carr:2016drx,Carr:2017jsz} for a review. The primordial black holes (dashed-dotted line) produced in Critical Higgs Inflation could comprise all of the dark matter and still pass current constraints. Note that a relatively narrow mass distribution of PBH does not change appreciably the constraints (dashed gray line).}
       \vspace{-10pt}
  \label{fig:consPBH}
\end{figure}

From equality to the present times, the mass distribution will shift to higher masses due to merging and accretion. In this CHI scenario, there is a very wide peak in the matter spectrum at small scales. This means that PBHs will cluster in very dense environments, which can significantly increase the frequency of black-hole-binary mergers~\cite{PBH}. In order to exactly determine the mass distribution of PBHs today, one would have to solve the non-linear evolution with a N-body simulation. Following Ref.~\cite{Chisholm}, we estimate the growth in PBH masses by a factor $3\times10^7$. In this case, we find that the peak of the lognormal distribution corresponds today to approximately $\mu_{_{\rm PBH}} \simeq 11\,\Msun$ and the lognormal dispersion to $\sigma_{_{\rm PBH}} \simeq 0.8$. Note that the mean of the PBH distribution is determined by the location of the maximum of the power spectrum $N_{\mathrm{peak}}$, leading to $\mu_{_{\rm PBH}}\sim 10\,\Msun\cdot e^{2(N_{\mathrm{peak}}-28.8)}$, while the variance is more sensitive to the width of the peak $\Delta N$. For the range we are considering, $\Delta N\in(30-35)$, then $\sigma_{_{\rm PBH}} \sim(0.6-1)$ and $N_{\mathrm{peak}}\sim(25-30)$. Therefore, Dark Matter is dominated today by PBH with masses in the range from $0.01$ to $100\,\Msun$. As a consequence, the CHI scenario is able to generate the high-mass black hole binary (BHB) mergers that have been observed by LIGO. Moreover, this mass distribution passes all observational constraints without difficulty, see Fig.~\ref{fig:consPBH}. Note that taking into account the non-zero width of the distribution, as in Ref.~\cite{Carr:2017jsz}, does not significantly change the constraints, since in our case $\sigma_{_{\rm PBH}}\simeq0.8$, and the peak of our mass distribution is well below the microlensing constraints. 

Apart from the direct GW emission from BHB mergers detected by LIGO, there are several stochastic backgrounds coming from different epochs. One GW background comes from unresolved BHB mergers since equality, with a power law spectrum, 
\be\nn
h^2\,\Omega_{\rm GW}(f) = 8\times10^{-15} \ \tau_m \,
f^{2/3}({\rm Hz})\,\mu^{5/3}(M_\odot)\,R(\sigma) \,, 
\ee
where $\tau_m\sim 50$ events/yr/Gpc$^3$ is the BHB merger rate and $R(\sigma)$ is an exponentially growing function of $\sigma$, see~\cite{GWB}. In the near future we may be able to detect this irreducible GW background with LISA~\cite{LISA}. A totally different background arises from second-order anisotropic stresses induced by large curvature fluctuations at horizon reentry when PBH formed, which has a broad peak in the sub nHz region and could eventually be detected by SKA~\cite{Garcia-Bellido:2017aan}.

\vspace{2mm}
{\it Conclusions}. In this letter we have explored the possibility that the Standard Model Higgs, with a non-minimal coupling to gravity, may have acted as the inflaton in the early universe, and produced all of the present dark matter from quantum fluctuations that reentered the horizon as huge curvature perturbations and collapsed to form black holes much before primordial nucleosynthesis. Taking into account the RGE running of both the Higgs self-coupling $\lambda$ and the non-minimal coupling to gravity $\xi$, we find regions of parameter space allowed by the Standard Model for which the inflaton-Higgs potential acquires a second plateau at smaller scales, around the critical point $\lambda(\mu) \simeq \beta_\lambda(\mu) = 0$. This plateau gives a super-slow-roll evolution of the Higgs, inducing a high peak in the curvature power spectrum which is very broad. When those fluctuations reenter the horizon during the radiation era they collapse to form primordial black holes with masses in the range $0.01$ to $100\,\Msun$, which could explain the LIGO events~\cite{LIGO}, and at the same time evade all of the present constraints on PBH~\cite{Carr:2016drx}. Some of these PBH may evaporate before equality; the rest will act as seeds for galactic structures~\cite{CGB} and initiate  reionization at high redshift~\cite{Kashlinsky:2016sdv}. Such a high peak in the matter power spectrum occurs at much smaller scales than are probed in present large scale structure surveys, but eventually its non-linear tails may be detectable in the future. Moreover, this scenario of massive PBH could explain the missing satellite problem, as well as the large mass-to-light ratios found in dwarf spheroidals~\cite{CGB,Li:2016utv}, and is consistent with Fermi-LAT gamma-ray observations~\cite{FermiLAT}. The stochastic background of gravitational waves from the merging of massive black hole binaries in the dense clusters after equality could be detectable by LISA or PTA~\cite{GWB,LISA}. Furthermore, this CHI scenario has also distinctive inflationary signatures, such as large curvature fluctuations at the end of inflation that may lead to a phase of inhomogeneous reheating. 

But, more importantly, the PBH-CHI scenario opens a new portal to test fundamental physics above the LHC scale. The RGE running of the SM Higgs couplings, from the electroweak scale to almost the Planck scale, may contribute to our understanding of the stability of the electroweak vacuum and, moreover, to constrain new physics beyond the Standard Model of Particle Physics.

\vspace{2mm}
{\it Acknowledgements}. JGB thanks Sebastian Clesse, Fedor Bezrukov, Misha Shaposhnikov and Javier Rubio for useful discussions. We also thank Yungui Gong for useful correspondence. The authors acknowledge support from the Research Project FPA2015-68048-03-3P (MINECO-FEDER) and the Centro de Excelencia Severo Ochoa Program SEV-2016-0597. JME is supported by the Spanish FPU Grant No. FPU14/01618. He thanks UC Berkeley and BCCP for hospitality during his stay there and UAM for financial support. ERM acknowledges support from the Programa Propio UPM. JGB acknowledges support from the Salvador de Madarriaga program Ref. PRX17/00056.

\end{document}